\documentclass[12pt]{article}
\begin{document}

\title{\bf \Large Force and Momentum in an Evolving Axisymmetric Universe Model}

\author{M. Sharif \thanks{e-mail: hasharif@yahoo.com}
\\ Department of Mathematics, University of the Punjab,\\ Quaid-e-Azam
Campus Lahore-54590, PAKISTAN.}

\date{}

\maketitle

\begin{abstract}
We take an axisymmetric rotating universe model by crossing with a
time dependent factor and evaluate its force and momentum in this
evolving universe. It is concluded that it behaves exactly like a
Friedmann model. We also extend this conclusion to the most
general cosmological model.
\end{abstract}

\section{Introduction}

There is no gravitational source in a Friedmann universe as it is
necessarily homogeneous. However, there are models which cross the
Friemann metric with a Schwarzschild metric [1], but only one
metric acts at any point in the spacetime. Thus, an observer can
either experience the expansion of the universe or the
gravitational attraction, but not both simultaneously. Bokhari and
Qadir [2] presented an alternative way of constructing a toy model
which gives an effect of a gravitational source in an evolving
universe. In this model, the spatial part of the metric is the
same as the Schwarzschild metric, but multiplied by a time
dependent scale factor $a(t)$. The time component is the usual
Schwarzschild time component. It was shown that the metric
considered do provide a Friedmann-like universe model with a
gravitational source present. Notice that this is merely a toy
model to deal with time evolution and gravitational force
simultaneously. We have extended this idea by constructing a new
metric [3,4] which is applicable to all the three Friedmann
models. It turned out that the force and the potential experienced
by a test particle are exactly the same as the Newtonian
gravitational force and potential respectively for each of the
Friedmann universe models.

In this paper we shall analyse an expanding axisymmetric
cosmological model with a rotating gravitational field by using
the same procedure. We shall restrict ourselves to seeing how this
model universe expands with time. It is concluded that the time
component of the force and the momentum exactly turn out to be
that of the Friemann universe. Further, we extend this conclusion
to the most general case.

To end this an operational procedure to express the consequences
of Relativity in terms of the Newtonian concept of gravitational
force has been used. The pseudo-Newtonian $(\psi N)$
``gravitational force'' is defined as the vector whose intrinsic
derivative along the separation vector is the maximum tidal force,
which is given by the acceleration vector for a preferred class of
observers. The proper time integral of the force four-vector will
be the momentum four-vector. We will not, here, discuss this
formalism rather we shall give the essential formula as it is
available elsewhere [5]. We shall use this formalism to calculate
the force and the momentum for the expanding axisymmetric universe
model.

In the free fall rest-frame the extended $\psi N(e\psi N)$ force
four-vector is given [3,6,7] by
\begin{eqnarray}
F_0=m[(\ln (Af))_{,0}+g^{ik}g_{jk,0}g^{jl}g_{il,0}/4A],\nonumber\\
F_i=m(\ln \sqrt {g_{00}})_{,i},\quad (i,j,k,l=1,2,3),
\end{eqnarray}
where $f=\frac{1}{\sqrt{g_{00}}}, A=(\ln \sqrt{-g})_{,0}$ and
$g=det(g_{ij})$. Thus the momentum four-vector, $p_\alpha$, is [7]
\begin{equation}
p_\alpha =\int F_\alpha dt, \quad (\alpha=0,1,2,3).
\end{equation}

\section{Non-Static Axisymmetric Cosmological Model}

We consider an expanding axisymmetric cosmological model whose
metric is given [8] by
\begin{equation}
ds^2=(e^\nu-e^\lambda\omega^2)dt^2-a^2(t)(e^\mu dr^2+e^\rho
d\theta^2+e^\lambda d\phi^2)+2a(t)e^\lambda\omega dtd\phi,
\end{equation}
where $\nu, \lambda, \mu$ and $\rho$ are functions of $r$ and
$\theta$ coordinates and $\omega$ is the constant angular velocity
of the gravitational field. It is mentioned here that there does
not exist any time-like Killing vector for this particular
universe model.

To be able to see the evolution of this universe model, we
evaluate force and momentum given by Eqs.(1) and (2). Using Eq.(3)
in Eq.(1), it follows that
\begin{equation}
F_0=m\frac{\ddot a}{\dot a},
\end{equation}
\begin{equation}
F_1=\frac{m(\nu^\prime e^{2\nu}-3\nu^\prime
e^{(\nu+\lambda)}\omega^2+2\lambda^\prime
e^{2\lambda}\omega^4)}{2(e^\nu-e^\lambda\omega^2)(e^\nu-2e^\lambda\omega^2)},
\end{equation}
\begin{equation}
F_2= \frac{m(\nu_\theta e^{2\nu}-3\nu_\theta
e^{(\nu+\lambda)}\omega^2+2\lambda_\theta
e^{2\lambda}\omega^4)}{2(e^\nu-e^\lambda\omega^2)(e^\nu-2e^\lambda\omega^2)},
\end{equation}
\begin{equation}
F_3=0.
\end{equation}
Using Eqs.(4)-(7) in Eq.(2), we obtain
\begin{equation}
p_0=m\ln (a_0\dot a),
\end{equation}
\begin{equation}
p_1=\frac{m(\nu^\prime e^{2\nu}-3\nu^\prime
e^{(\nu+\lambda)}\omega^2+2\lambda^\prime
e^{2\lambda}\omega^4)}{2(e^\nu-e^\lambda\omega^2)(e^\nu-2e^\lambda\omega^2)}t+a_1,
\end{equation}
\begin{equation}
p_2=\frac{m(\nu_\theta e^{2\nu}-3\nu_\theta
e^{(\nu+\lambda)}\omega^2+2\lambda_\theta
e^{2\lambda}\omega^4)}{2(e^\nu-e^\lambda\omega^2)(e^\nu-2e^\lambda\omega^2)}t+a_2,
\end{equation}
\begin{equation}
p_3=a_3.
\end{equation}
where $a_0, a_1, a_2, a_3$ are integration constants. The dot,
prime and $\nu_\theta$ denote differentiation with respect to
time, $r$ and $\theta$ respectively. We see that the temporal
component of the force four-vector is exactly the same as the
Friedmann model. The spatial components of the force are in
agreement with the already evaluated force for the axisymmetric
metric. This indicates that the metric given by Eq.(3) infact
yields the effect of a rotating gravitational field. Further, the
time component of the momentum 4-vector is the same as for the
Friedmann universe model and its analysis can be seen [9].

\section{Extension to the most general cosmological model}

Notice that the temporal component of the force four-vector has
time derivative only. The first term contains $\ln
\frac{A}{\sqrt{g_{00}}}$ which, due to ``$\ln$'' property,
separates $\sqrt{g_{00}}$ and so the time derivative of this
$g_{00}$ term always vanishes as it remains independent of time
even in the most general case. The second term contains terms like
$g_{ij}$ which involve scale factor $a(t)$. Thus it is easy to see
from here that we have only scale factor $a(t)$ like terms in
$F_0$ which always give the Friedmann like result. The spatial
components $F_i$ has space coordinate derivative and so it would
always yield the usual pseudo-Newtonian force.  Now we show this
result by taking a most general model.

The most general cosmological model (a toy model) can be
constructed by keeping the spatial part of the metric same, but
multiplied by a time dependent scale factor. The time component is
the usual one. It is given (as every metric can be diagonalised by
using a suitable coordinate transformations [3,10]) by
\begin{equation}
ds^2=e^{2\lambda}dt^2-a^2(t)[e^{2\mu}dr^2+e^{2\nu}d\Omega^2],
\end{equation}
where $\lambda,\mu,\nu$ are arbitrary functions of $r,\theta,\phi$
and $d\Omega^2=d\theta^2+\sin^2\theta d\phi$ is the solid angle.

Now we evaluate four-vector force and momentum for this hand
constructed toy cosmological model. Inserting all the terms
involving in Eqs.(1) and (2) from Eq.(12), it follows that
\begin{equation}
F_0=m\frac{\ddot a}{\dot a},\quad F_1=m\lambda',\quad
F_2=m\lambda_\theta,\quad F_3=m\lambda_\phi,
\end{equation}
where $\lambda',\lambda_\theta$ and $\lambda_\phi$ denote
differentiation with respect to the coordinates $r,\theta$ and
$\phi$ respectively. Substituting Eq.(13) in Eq.(2), we have
\begin{equation}
p_0=m\ln (a_0\dot a),\quad p_1=m\lambda't+a_1,\quad
p_2=m\lambda_\theta t+a_2,\quad p_3=m\lambda_\phi t+a_3.
\end{equation}
It follows from Eq.(13) that the temporal component of the force
four-vector is exactly the same as the Friedmann model while the
spatial components of the force are in agreement with the already
evaluated force for the general metric. Again, we see that the
time component of the momentum 4-vector is the same as for the
Friedmann cosmological model.

\section{Conclusion}

A non-static axisymmetric cosmological model universe is
considered with a rotating gravitational field. The four-vector
force and momentum have been evaluated for this model universe.
This shows that the temporal component of the force and momentum
four vector turn out exactly the same as for the Friedmann
cosmological model while the spatial components are the usual one.

Further, we have extended this idea to the most general case by
constructing a hand made cosmological model and evaluated its
four-vector force and momentum. We note that the temporal
component of the four-vector force and momentum always remain the
same as for the Friedmann model. Thus we conclude that whatever
cosmological model we construct, by crossing with a time dependent
scale factor $a(t)$, it always provide a Friedmann-like universe
behaviour.

\newpage\

\begin{description}
\item {\bf Acknowledgments}
\end{description}

The author would like to thank Prof. Chul H. Lee for his
hospitality at the Department of Physics and Korea Scientific and
Engineering Foundation (KOSEF) for postdoc fellowship at Hanyang
University Seoul, KOREA.

\vspace{2cm}

{\bf \large References}

\begin{description}

\item{[1]} Lindquist R.W. and Wheeler J.A. Rev. Mod. Phys. {\bf 29}(1957)432;\\
Einstein A. and Strauss E. Rev. Mod. Phys. {\bf 17}(1945)120; {\bf 18}(1946)148.

\item{[2]} Bokhari A.H. Ph.D. Thesis Quaid-i-Azam University Islamabad (1985);
Bokhari A.H. and Qadir A. {\it Proc. 4th Marcel Grossmann Meeting
on General Relativity}, ed. R. Ruffini (Elsevier Science
Publishers 1986) 1635.

\item{[3]} Sharif M. Ph.D. Thesis Quaid-i-Azam University Islamabad (1991).

\item{[4]} Sharif M. {\it Effects of Force and Energy in an Evolving Universe with a Gravitational
Source} submitted for publication.

\item{[5]} Mahajan M., Qadir A. and Valanju P.M. Nuovo Cimento B {\bf 65}(1981)404;
Qadir A. and Quamar J. {\it Proc. 3rd Marcel Grossmann Meeting on
General Relativity}, ed. Hu Ning (North Holland Amstderm
1983)189;\\ Qaumar J. Ph.D. Thesis Quaid-i-Azam University
Islamabad (1984).

\item{[6]} Qadir Asghar and Sharif M. Nuovo Cimento B {\bf 107}(1992)1071.

\item{[7]} Qadir Asghar and Sharif M. Phys. Lett. A {\bf 167}(1992)331;\\
Sharif M. Astrophys. And Space Science {\bf 253}(1997)195.

\item{[8]} Tiwari R.N., Rao J.R. and Kanakamedala R.R. Phy. Rev. D {\bf 34} (1986) 327;\\
Bokhari A.H. Int. J. Theor. Phys. {\bf29}(1990)1011.

\item{[9]} Sharif M. Astrophysics and Space Science {\bf 262}(1999)297-304.

\item{[10]} Landau L.D. and Lifschitz E.M.: {\it The Classical
Theory of Fields} (Pergamon Press, 1975).

\end{description}

\end{document}